\documentclass[a4paper,reqno,twoside]{amsart}
\usepackage[latin1]{inputenc}
\usepackage[T1]{fontenc}
\usepackage{textcomp}
\usepackage[british]{babel}
\usepackage{newcent} 
\usepackage[dvips,pagebackref]{hyperref}
\usepackage{xspace}
\RequirePackage{mathrsfs} 
\usepackage{amsmath}
\usepackage{amssymb}
\usepackage{amsthm}
\usepackage{amsfonts}
\newcommand{\CC}{{\ensuremath{\mathord{\mathbb{C}}}}\xspace}
\newcommand{\RR}{{\ensuremath{\mathord{\mathbb{R}}}}\xspace}

\newcommand{\NN}{{\ensuremath{\mathord{\mathbb{N}}}}\xspace}

\newcommand{\ZZ}{{\ensuremath{\mathord{\mathbb{Z}}}}\xspace}
\DeclareMathAlphabet{\Bmi}{OT1}{cmm}{b}{it}
\newcommand{\EPS}{\ensuremath{\varepsilon}\xspace}

\newcommand{\RE}{\operatorname{Re}}
\newcommand{\supp}{\operatorname{supp}}

\newcommand{\sign}{\operatorname{sign}}
\newcommand{\diag}{\operatorname{diag}}

\newcommand{\ABS}[1]{{\ensuremath{\mathord{\left|#1\right|}}}}
\newcommand{\NORM}[1]{{\mathord{\left\|#1\right\|}}}
\newcommand{\DEF}{\mathbin{\smash[t]{\overset{\scriptscriptstyle\mathrm{def}}{=}}}}

\newcommand{\SPROD}[2]{
            \ensuremath{\mathord{\left\langle#1\,\mathord{,}\,#2\right\rangle}}\xspace}
\newcommand{\IPROD}[2]{
            \ensuremath{\mathord{\left(#1\,\mathord{,}\,#2\right)}}\xspace}
\newcommand{\LIN}[1]{
            \ensuremath{\mathord{\left\langle#1\right\rangle}}\xspace}
\newcommand{\dd}{\ensuremath{\mathrm{d}}} 
\newcommand{\ee}{\ensuremath{\mathrm{e}}} 
\newcommand{\ie}{, {\it i.e.},\xspace}
\newcommand{\eg}{, {\it e.g.},\xspace}
\newcommand{\GLQQ}{\char"12\kern .08em} 
\newcommand{\GRQQ}{\kern .08em\char"10\xspace}
\def\SetConstructor#1#2#3#4{%
  \def\test{#4}\ifx\test\empty{\ensuremath{\mathord{#1}_{#2}^{#3}}\xspace}%
  \else{\ensuremath{\mathord{#1}_{#2}^{#3}(#4)}\xspace}\fi}
\newcommand{\SN}{\ensuremath{\mathord{\mathscr{S}}}\xspace}    
\newcommand{\SIDX}[3]{\SetConstructor{\mathscr{S}}{#1}{#2}{#3}}
\newcommand{\SUD}[3][]{\SIDX{#3}{#2}{#1}}                      
\theoremstyle{plain}
  \newtheorem{theo}{Theorem}[section]
\newtheorem*{theo*}{Theorem}

\newtheorem{prop}[theo]{Proposition}
\newtheorem{coro}[theo]{Corollary}
\newtheorem{lemm}[theo]{Lemma}
\theoremstyle{definition}
\newtheorem{defi}[theo]{Definition}
\newtheorem{defirem}[theo]{Definition and Remark}

\renewcommand{\labelenumi}{\roman{enumi})}
\theoremstyle{remark}

\newcommand{\labelS}[1]{\label{sect:#1}}
\newcommand{\labelT}[1]{\label{theo:#1}}
\newcommand{\labelP}[1]{\label{prop:#1}}
\newcommand{\labelL}[1]{\label{lemm:#1}}
\newcommand{\labelD}[1]{\label{defi:#1}}
\newcommand{\labelC}[1]{\label{coro:#1}}
\newcommand{\refC}[1]{Corollary~\ref{coro:#1}}
\newcommand{\refS}[1]{Section~\ref{sect:#1}}
\newcommand{\refT}[1]{Theorem~\ref{theo:#1}}
\newcommand{\refP}[1]{Proposition~\ref{prop:#1}}
\newcommand{\refL}[1]{Lemma~\ref{lemm:#1}}
\newcommand{\refD}[1]{Definition~\ref{defi:#1}}
\newcommand{\refE}[1]{equation~\eqref{eq:#1}}
\newif\ifpdf
\pdftrue
\newcommand{\SV}{\ensuremath{\mathord{\mathcal{V}}}\xspace}
\newcommand{\SP}{\ensuremath{\mathord{\mathcal{P}}}\xspace}
\newcommand{\FOUR}{\ensuremath{\mathord{\mathcal{F}}}\xspace}
\newcommand{\SH}{\ensuremath{\mathord{\mathcal{H}}}\xspace}
\newcommand{\SK}{\ensuremath{\mathord{\mathcal{K}}}\xspace}
\newcommand{\SNN}{\ensuremath{\mathord{\mathcal{N}}}\xspace}
\newcommand{\MOM}[2][]{\SetConstructor{\mu}{}{#2}{#1}}
\begin{document}
%
%
\title[Infinite Infrared Regularization]{Infinite 
Infrared Regularization\\
and a State Space for the Heisenberg Algebra}
\author[A.\ U.\ Schmidt]{Andreas U.\ Schmidt}
\date{20th February 2002}
\curraddr{Dipartimento di Fisica E.\ Fermi\\
  Universit\`a di Pisa\\
  via Buonarroti 2, Ed.\ B\\
  56127 Pisa\\
  Italy}
\address{Fachbereich Mathematik\\
  Johann Wolfgang Goethe-Universität\\
  60054 Frankfurt am Main, Germany\ifpdf\\
  \href{http://www.math.uni-frankfurt.de/~aschmidt}{Homepage}\else\fi}
\ifpdf
  \email{\href{mailto:aschmidt@math.uni-frankfurt.de}{aschmidt@math.uni-frankfurt.de}}
\else
 \email{aschmidt@math.uni-frankfurt.de}
\fi
\thanks{This research was supported by a research grant from the
Deutsche Forschungsgemeinschaft 
\ifpdf\href{http://www.dfg.de}{DFG}\else DFG\fi. The author wishes to thank 
the University of Durban--Westville for its hospitality and 
South Africa for the weather.
Heartfelt
thanks go to Erwin Br\"uning (Durban), Giovanni Morchio, 
and Franco Strocchi (both Pisa), 
Daniel Lenz, and Matthias Schork (both Frankfurt am Main) for
many useful discussions.}
\begin{abstract}
We present a method for the construction of a Krein space completion
for spaces of test functions, equipped with an indefinite inner product 
induced by a kernel which is more singular than a distribution
of finite order. This generalizes a regularization method
for infrared singularities in quantum field theory, introduced by 
G.~Morchio and F.~Strocchi, to the case of singularites of infinite order.
We give conditions for the possibility of this procedure in terms
of local differential operators and the Gelfand-Shilov test function spaces,
as well as an abstract sufficient condition. 
As a model case we construct a maximally positive definite state space
for the Heisenberg algebra in the presence of an infinite infrared singularity.
\end{abstract}
\maketitle
%
%
\section{Introduction}
A notable case in which some of the abundant singularities of quantum field theory
can be treated rigorously is presented by
the method of infrared regularization of Morchio and Strocchi~\cite{MPS90,b:STR93}.
There, the first-order singularity of the two-point function of the massless scalar
field in $1+1$-dimensional spacetime manifests itself in the nonpositivity
of the Wightman inner product on the one-particle space. 
In momentum space, this two-point function appears
as a singular integral kernel which is regularized in the distributional sense
as a Cauchy principal value. Since this regularization involves subtraction of
values of the test functions at $p=0$, the Wightman inner product induced
by the two-point function is clearly no longer positive definite. It turns out that,
if the usual positivity axiom of Wightman theory, see~\cite{b:SW64}, is
replaced by a weaker Hilbert space structure condition, the construction of
a suitable physical state space is still possible. 
The one-particle space becomes a Krein
space, the natural analogon of a Hilbert space in the case of an indefinite
inner product, and it is maximal in the sense that there is no larger Krein
space closure of the test function space (we refer to Appendix~\ref{sect:indef},
where some basic notions of, and results on, indefinite inner product spaces are
gathered). Thus, no physical information gets lost and one can identify a 
positive definite physical Hilbert subspace. In fact, in the case treated 
in~\cite{MPS90,b:STR93} the rank of negativity is one, and thus the Krein
space is actually a Pontryagin space. The basic principles of this regularization
procedure in the rank one case have already been noted in~\cite{DT66}
(The author wishes to thank Daniel Dubin for bringing this reference
to his attention).

In~\cite{SCH97A}, we have cast this procedure
in abstract form, yielding a method by which every quasi-positive space\ie
a space with finite rank of negativity, can be completed to a Pontryagin space.
By this, we generalized the infrared regularization method to singularities of the type
of finite order distributions. On the other hand it is by now well known that
constructive approaches to interacting quantum fields generically involve much more
singular objects, see~\cite{WIG81,MS92}, namely ultradistributions and even
Fourier Hyperfunctions~\cite{BN89,BN98}. Thus, it is natural to look for
a further generalization of the procedure for finding a maximal Krein space closure
starting from a space of test functions with an indefinite inner
product induced by a singular kernel, to the case of non-distributional
infnite order singularities, and therefore to the case of an infinite number of negative
degrees of freedom. This is what we will present in the following.

We illustrate the regularization method\footnote{We tend to denote the whole process
of defining an indefinite inner product by a generalized function \emph{and} the
construction of a maximal Krein space closure as `regularization'. The first step,
which is the traditional regularization of a singular integral, means going
only half way toward a physically conclusive result.} 
by a neat (yet unphysical) model
in which infinite order singularities appear naturally, at least on a 
heuristic level. Namely, we will consider the Schr\"odinger representation
of the Heisenberg algebra on a test function space over \RR in the presence
of a singularity concentrated at $p=0$. This model will be informally described
in the next section, where also some notions needed subsequently are introduced.
Further, we will state the main result, which is that the
regularization procedure yields a maximal Krein space and in it a 
largest possible, positive definite, closed subspace, 
which can sensibly be considered as the `physical 
state space' for the Heisenberg algebra of `observables'.

\refS{regularization} contains the regularization procedure proper. It shows
in particular that there is a certain balance that has to be kept between the
singularity of the inner product, measured in terms of  infinite order (local)
differential operators, and the choice of test function space, which we
express in terms of the Gelfand-Shilov scheme of spaces, see~\cite[Chapter~IV]{b:GS64}.
The method itself is, however, general enough to be applied to a much wider
class of singularities than infrared ones, and for a lot of other test function
spaces.

In \refS{generalized}, we formulate abstractly a sufficient condition
under which the regularization is guaranteed to work. The conditions we give 
are not the most general and  abstract ones possible, since they reflect the
limitations of the procedure of \refS{regularization}. Therefore they present 
no sharp criterion to decide whether
an indefinite space admits the construction of a maximal majorant topology
by our construction of a Hilbert majorant. Nevertheless, they
capture the essential points that enable our construction, and therefore 
are at least useful to explain the mechanism behind it. 
Furthermore, our conditions are simple enough to be effective in many concrete
cases. We discuss possible further generalizations at the end of \refS{generalized}.

Appendix~\ref{sect:neutral} contains a simple, concrete construction of certain
neutral elements\ie vectors with vanishing inner product, which play an important
role in the regularization procedure in \refS{regularization}. 
Appendix~\ref{sect:indef} compiles some
basics about indefinite inner product spaces mainly taken from~\cite{b:BOG74}.
\section{The Model}
Recently, a comprehensive abstract classification of representations 
of the Hei\-senberg algebra on an indefinite inner product space has 
been worked out by Mnatsakanova, Morchio, Strocchi, and Vernov in~\cite{MMSV98}.
There, it was pointed out that this issue is somewhat more difficult to handle
than in the positive definite case, which is covered by the Stone-von Neumann 
uniqueness theorem, see~\cite[Chapter IV]{b:PUT67}. Especially, domain questions 
appear and the notion of irreducibility has to be reconsidered. Here, we take 
a different approach in considering a very concrete example where the Heisenberg
representation is from the beginning assumed to be the quantum mechanical Schr\"odinger
representation
\[
\hat{q}\DEF x\cdot, \quad \hat{p}\DEF-i\frac{\dd }{\dd x}\;, 
\]
on a function space over \RR.

The following discussion will take place in momentum space, and we will 
notoriously denote the Fourier transforms of functions with $f$, $g$, etc.,
and the variable by $p$. Consider the indefinite inner product
\begin{equation}
\label{eq:iprod}
\SPROD{f}{g}\DEF\IPROD{f}{g}_{L^2}-\sum_{k=0}^\infty c_k^2
\overline{f}^{(k)}(0)g^{(k)}(0),
\end{equation}
with real coefficients $c_k$ 
(this will turn out to pose no essential restriction in our case, see the following). 
It can be formally interpreted as being
induced by a generalized function (a kernel) on $\RR^2$ in the following way:
\begin{equation}
\SPROD{f}{g}=\IPROD{\delta(p-p')-
J(\partial_{p}\partial_{p'})\delta(p)\delta(p')}{\overline{f}(p)g(p')}.
\label{eq:distribution}
\end{equation}
Here, the infinite order differential operator $J$ is given by its
symbol
\[
J(\xi)\DEF\sum_{k=0}^\infty c_k^2 \xi^k.
\]
The singularity in \SPROD{.}{.} can be characterized by the 
following notion, where we have already adapted the conventional 
notation a bit, so as to conform with our application in \refS{regularization}:
\begin{defirem}[{\cite[159 pp.]{b:GS64}}]
  An entire function $J(\xi)$ in the complex variable $\xi$ is called
\textit{infra-exponen\-ti\-al of order $1/(2\delta)$} if it fulfills
for every $\EPS>0$ an estimate
\[
\ABS{J(\xi)}\leq C_{\EPS} \ee^{\EPS\ABS{\xi}^{1/(2\delta)}},
\]
for some $C_{\EPS}>0$. In this case, the coefficients $c_k^2$  of the
 Taylor series of $J$ satisfy the following
upper bounds: For every $D>0$ exists a $\theta\in(0,1)$ and a $C>0$
such that
\begin{equation}
  \label{eq:singularity}
  \ABS{\smash{c_k^2}}\leq C\frac{\theta^k}{D^k\ee^{2k\delta}k^{2k\delta}}.
\end{equation}
\end{defirem}
Now, our first concern is on which test function
space the inner product can be defined. 
To this end, we use the Gelfand-Shilov scheme for
the classification of spaces of smooth functions, see~\cite[Chapter IV]{b:GS64}. 
For $0\leq\alpha, \beta\leq\infty$
the space \SUD[\RR]{\beta}{\alpha} consists of smooth functions  
$f$ on \RR satisfying estimates
\[
\ABS{\smash{p^qf^{(k)}(p)}}\leq C A^qB^k q^{q\alpha}k^{k\beta}.
\]
We need in fact only consider the regularity of the functions in 
\SUD{\beta}{\alpha} at the origin, which
is expressed in the following basic estimate: There exists a $B>0$ such that
for all $\rho>0$ and a constant $C_f$ depending on $f$ we have
\begin{equation}
  \label{eq:regularity}
  \ABS{\smash{f^{(k)}(0)}}\leq C_f (B+\rho)^k k^{k\beta}.
\end{equation}
It is apparent from~\eqref{eq:singularity} and~\eqref{eq:regularity} that
the indefinite inner product is well-defined on \SUD{\beta}{\alpha} by~\eqref{eq:iprod}
as the distribution~\eqref{eq:distribution}, whenever
$J$ is infra-exponential of order $\leq1/(2\beta)$.
We denote by $\SV=\SV_J(\alpha,\beta;\rho)$ a space 
\SUD{\beta}{\alpha} equipped with an indefinite
inner product~\eqref{eq:iprod} defined by an infra-exponential symbol $J$ 
of order $1/(2\delta)$ for any $\delta\leq\beta$. 

Further constraints on the choice of test function space now come from the
intended Schr\"odinger representation of the Heisenberg algebra. 
As the Heisenberg generators act 
in momentum space by multiplication with $p$ and differentiation $i\dd/\dd p$, it
is clear that they will not be symmetric operators with respect to the indefinite 
inner product \SPROD{.}{.} on the whole space. 
A representation on a subspace of \SV acts by
symmetric operators only if this subspace consists of functions $f$ such that
all derivatives of $f$ vanish at $p=0$. This subspace is at the same time
also positive definite. In order that it is also maximal in the sense that
there is no larger positive definite subspace in \SV, one has in fact to assume
that the coefficients $c_k^2$ in \refE{iprod} are strictly positive\ie 
$c_k\in\RR\setminus\{0\}$ for all $k$, which we do from now on. 
In turn, this implies that 
$J(\partial_p\partial_{p'})$ is a properly infinite differential operator and
thus the singularity in~\eqref{eq:distribution} must be stronger than a finite
order distribution. This excludes as test function space any space 
\SUD{\infty}{\alpha} which allows only distributions of finite order, 
and thus in particular the Schwartz space $\SN=\SUD{\infty}{\infty}$.
On the other hand, the very strong singularity of an analytic functional
is also excluded: Since for $\beta=1$ the test functions in \SV are all analytic
in a strip neighbourhood of the real axis, the requirement $f^{(k)}(0)=0$ for all
$k$ would lead to the trivial subspace. 

After these heuristics, we are ready to
state our main result, whose proof will follow in the next section.
We will characterize the complete, positive definite representation
subspace for the Heisenberg algebra by the Fourier transformation.
For that, we need another definition. 
\begin{defi}
\labelD{Lnull}
The space of functions $L_0^2(\RR)$ is defined as
\[
L_0^2(\RR)\DEF\Bigl\{ f\in L^2(\RR) \Bigm| \MOM[f]{k}=0,\ \forall k\in\NN_0\Bigr\},
\]
where the  \textit{$k$th moment} \MOM[f]{k} of a function $f\in L^2(\RR)$ 
is given by
\[
\MOM[f]{k}\DEF\int_\RR x^k f(x) \dd x,
\] 
if it exists for a $k\in\NN_0$.
\end{defi}
\begin{theo}
  \labelT{main}
  Let $0\leq\alpha\leq\infty$, $1<\beta<\infty$. 
Assume $\delta>\beta$, and let $J$ be infra-exponential of order 
$1/(2\delta)$ with strictly positive Taylor coefficients. 
Then, the space $\SV=\SV_J(\alpha,\beta;(2\delta)^{-1})$ admits a 
maximal completion to 
a Krein space \SK with countably infinite rank of negativity. 
The maximal positive definite subspace of \SK which 
is invariant under the action of the Heisenberg algebra in the Schr\"o\-dinger 
representation by selfadjoint operators on it, is the Fourier
transform $\FOUR L_0^2(\RR)$ of $L_0^2(\RR)$.
\end{theo}
The appearance of $\delta>\beta$ results from technicalities of the
infrared regularization process, as will become clear in the following. This 
leaves room for improvement. It should be stressed that the diagonal
form $\delta(p-p')$  of~\eqref{eq:distribution} 
outside the singularity $p=p'=0$ was chosen
to allow for a symmetric action of the Heisenberg generators. The
regularization procedure itself is nonetheless rather independent 
of the structure of the kernel outside the singular points. On the other hand,
the discussion in the beginning of this section also points to
a principal limitation of the regularization method. If the singularity
in a certain point $p_0$ is that of a proper analytic functional\ie $\delta\leq1$,
and the rank of indefiniteness is infinite, regularization is impossible
since the positive subspace would be trivial in that case.

The somewhat exotic representation of the Heisenberg algebra above does not
fit into the classification of~\cite{MMSV98}, see also~\cite{MS00A}. 
Rather it corresponds to
the `counterexample' in the appendix of~\cite{MMSV98}. As explained there,
$L^2_0(\RR)$ naturally decomposes into two irreducible subspaces of
`left-' and `right-movers'\ie states with support on the negative, respectively 
positive half-axis in momentum space, by closure of the domains
$\SV_\pm=\{f\in\SV\mid f(p)=0\text{ for } p\lessgtr 0\}$ in the Krein topology.
\section{Infinite Infrared Regularization}
\labelS{regularization}
In this section we present the proper method for the construction of a Krein
space from \SV. The general strategy is close in spirit to the well-known
method for closing a Hilbert space with respect to the action
of a given positive bilinear form on it, see~\cite[Appendix~A.2]{b:GJ87}.
The construction of a maximal majorant Hilbert topology for \SV leading
to a Krein space closure of it, relies mainly on two ingredients: First,
the existence of neutral elements within \SV which separate the negative 
degrees of freedom from the rest of the space. Second, there is some
`air' left between the decay of the coefficients $c_k^2$ defining \SPROD{.}{.} 
via~\eqref{eq:iprod}, and the growth of the Taylor coefficients of the
functions in \SV. This is expressed in \refE{singularity}, the assumption
$\delta>\beta$ of \refT{main}, and~\eqref{eq:regularity}. To make use of that
margin, we define the `damping coefficients' $\gamma_k$ by
\begin{equation}
  \label{eq:damping}
  \gamma_k\DEF k^{k\delta}.
\end{equation}
The neutral decomposition elements will be constructed in 
Appendix~\ref{sect:neutral} to fulfill the following demands:
\begin{lemm}
\label{lemm:decomposing_functions}
  Let $0\leq\alpha\leq\infty$ and $1<\beta<\infty$. Let there
be given a sequence of numbers $c_k$ satisfying~\eqref{eq:singularity},
and let $\gamma_k$ be as in~\eqref{eq:damping}. 
Then there exists a sequence of functions 
$\{\chi_k\}_{k\in\NN_0}\subset\SUD{\beta}{\alpha}$ with the following
properties:
\begin{enumerate}
\item $\NORM{\chi_k}_{L^2}^2=c_k^2\gamma_k^2$.
\item $\chi_k^{(i)}(0)=\delta_{ik}\cdot\gamma_k$.
\item $\IPROD{\chi_k}{\chi_l}_{L^2}=0$, for all $k\neq l$.
\item $\SPROD{\chi_k}{\chi_l}=0$, for all $k$, $l$. 
\end{enumerate}
We denote by \SNN the linear subspace of \SUD{\beta}{\alpha} generated
by $\{\chi_k\}_{k\in\NN_0}$.
\end{lemm}
The subspace \SNN is neutral, $\SNN\subset\SV^0$. 
We also observe that \SV is non-de\-ge\-ne\-ra\-te, due to the presence of
the $L^2$-part in the indefinite product~\eqref{eq:iprod}. This property 
will prevail in the closure of \SV we construct in the following.
Now, every $f\in\SUD{\beta}{\alpha}$ has, for every finite $N\geq0$, a 
unique decomposition
\begin{equation}
f=f^{N+}+\sum_{i=0}^N f^i\chi_i,
\quad\text{with }f^i=\frac{f^{(i)}(0)}{\gamma_i},
\label{eq:finite_decomp}
\end{equation}
and $f^{N+}\in\SUD{\beta}{\alpha}$ is such that $f^{(i)}(0)=0$ for $0\leq i \leq N$.
Furthermore, the sum in the decomposition is clearly in \SNN.
\begin{prop}
\labelP{majorant}
  The seminorm $p$ given by the limit
  \begin{equation}
    \label{eq:majorant}
    p(f)^2\DEF
    \lim_{N\rightarrow\infty}\left[
    \SPROD{\smash{f^{N+}}}{\smash{f^{N+}}}+
    \sum_{i=0}^N \left\{
    \ABS{\SPROD{f}{\chi_i}}^2 + \ABS{\smash{f^i}}^2\right\}
    \right]
  \end{equation}
exists and defines a majorant topology $\tau$ on \SV.
\end{prop}
\begin{proof}
Taking \refL{norm_maj} into account, we have to show that 
\eqref{eq:majorant}, if it is well defined, dominates the inner square. 
Assuming that the limit in question exists, it is easy to show
that $p(f)^2$ majorizes the inner square $\ABS{\SPROD f f }$ of $f$. 
Namely, using~\eqref{eq:finite_decomp} we can express \SPROD{f}{f} as
\[
\SPROD{f}{f}=\SPROD{\smash{f^{N+}}}{\smash{f^{N+}}}+
\sum_{i=0}^N \left\{
f^i\SPROD{f}{\chi_i} + \overline{f^i}\SPROD{\chi_i}{f}\right\}
\]
using property iv) of~\refL{decomposing_functions}, and the fact that
$\SPROD{\smash{f^{N+}}}{\chi_i}=\SPROD{f}{\chi_i}$ which follows from it.
Now, in every term in the sum above we have the elementary
estimate for complex numbers
$\ABS{\smash{f^i\SPROD{f}{\chi_i} + \overline{f^i}\SPROD{\chi_i}{f}}}\leq
\ABS{\SPROD{f}{\chi_i}}^2+\ABS{\smash{f^i}}^2$. 
If the first term \SPROD{\smash{f^{N+}}}{\smash{f^{N+}}}
in~\eqref{eq:majorant} has a limit at all, then it tends to
$\IPROD{f^+}{f^+}_{L^2}=\NORM{f^+}_{L^2}^2\geq0$ for a certain $f^+\in L^2$,
showing $p(f)^2\geq\ABS{\SPROD f f }$ in the limit $N\rightarrow\infty$.
It remains to show that all the limits involved
in~\eqref{eq:majorant} exist. In order to show finiteness of 
the first term  it suffices to show that the 
decomposition~\eqref{eq:finite_decomp} of $f$ converges in $L^2(\RR)$ for 
$N\rightarrow\infty$, 
since it then tends to
$\NORM{f^+}_{L^2}^2$ and is thus necessarily finite as we have just seen. 
For the sum defining the decomposition, we have by i) of~\refL{decomposing_functions}
\[
\NORM{\smash{\sum_{N=0}^\infty} f^i\chi_i}_{L^2}^2\leq
\sum_{N=0}^\infty\NORM{\smash{f^i\chi_i}}_{L^2}^2=
\sum_{N=0}^\infty\ABS{\smash{f^{(i)}(0)c_i}}^2<\infty,
\]
taking~\eqref{eq:singularity} and~\eqref{eq:regularity} into account, showing
that claim. By definition~\eqref{eq:iprod} of the inner product, the $i$th 
term in the sum in~\eqref{eq:majorant} becomes
\begin{align*}
\ABS{\SPROD{f}{\chi_i}}^2 + \ABS{\smash{f^i}}^2&=
\Bigl|\IPROD{f}{\chi_i}_{L^2}-
\sum_{k=0}^\infty c_k^2 \overline{f}^{(k)}(0)\chi_i^{(k)}(0)\Bigr|^2 + 
\ABS{\smash{f^i}}^2\\
&\leq \ABS{\smash{\IPROD{f}{\chi_i}_{L^2}}}^2 +
c_i^4 \gamma_i^2 \ABS{\smash{f^{(i)}(0)}}^2 + 
\frac{ \ABS{\smash{f^{(i)}(0)}}^2}{\gamma_i^2}.
\end{align*}
For the first term we find, using the Cauchy--Schwartz estimate,  
by~\eqref{eq:singularity} and~\eqref{eq:damping}, and of course 
the assumption of \refT{main}:
\[
\ABS{\smash{\IPROD{f}{\chi_i}_{L^2}}}^2\leq
\NORM{f}_{L^2}^2\ABS{\smash{c_i^2\gamma_i^2}}\leq
\NORM{f}_{L^2}^2 \frac{C\theta^{i}}{D^{i}\ee^{2i\delta}}.
\]
Further using~\eqref{eq:regularity}, the second term is bounded by
\[
c_i^4 \gamma_i^2 \ABS{\smash{f^{(i)}(0)}}^2\leq
\frac{C^2C_f^2}{i^{2i(\delta-\beta)}} 
\left(\frac{\theta(B+\rho)}{\ee^{2\delta}D}\right)^{2i}.
\]
Finally the third term satisfies
\[
\frac{ \ABS{\smash{f^{(i)}(0)}}^2}{\gamma_i^2}\leq
\frac{C_f^2}{i^{2i(\delta-\beta)}}(B+\rho)^{2i}.
\]
All three terms decay faster than exponentially in $i$, making the 
overall sum in~\eqref{eq:majorant} convergent in the limit 
$N\rightarrow\infty$.
\end{proof}
Notice
that although we chose to see this independently by considering $L^2$-convergence, 
the numerical convergence of the $\SPROD{\smash{f^{N+}}}{\smash{f^{N+}}}$-part
of the decomposition could have been inferred in the same way as the convergence of the
other terms in~\eqref{eq:majorant}. In fact, one could have
inverted the decomposition~\eqref{eq:finite_decomp}
to yield $f^{N+}=f-\sum_{i=0}^Nf^i\chi_i$ and then see the convergence of 
$\SPROD{\smash{f^{N+}}}{\smash{f^{N+}}}$ by majorizing  it with the same
convergent terms as in the previous proof. 
\label{conv_note} 

It is apparent from the proof of~\refP{majorant} that the 
decomposition~\eqref{eq:finite_decomp} of $f$
\newcommand{\CLOT}[1]{\ensuremath{\mathord{\overline{#1}^\tau}}\xspace}
converges in the closure \CLOT{\SV} of
\SV with respect to $\tau$. In fact, it is easy to see that
the increments $p(f^i\chi_i)^2$ decay fast enough to turn the partial sums in the
decomposition into a Cauchy sequence. This allows us to write for
every $f\in\SV$,
\begin{equation}
f=f^{+}+\sum_{i=0}^\infty f^i\chi_i,
\quad\text{with }f^+\in\overline{\SV}^\tau.
\label{eq:decomposition}
\end{equation}
We further see, using the joint continuity of \SPROD{.}{.}, see~\refD{majorant},
that the indefinite inner product \SPROD{.}{.} has a
unique extension to \CLOT{\SV}, which we will also denote by \SPROD{.}{.}.
Thus using~\eqref{eq:decomposition}, \refE{majorant} extends to a definition 
of a quadratic normed topology on the $\tau$-complete space $\CLOT{\SV}$\ie a
Hilbert majorant topology on that space:
\begin{coro}
\labelC{Hilbert}
  On the closure $\SK\DEF\CLOT{\SV}=\CLOT{\SUD{\beta}{\alpha}}$  
we define the Hilbert scalar product
\begin{equation}
\label{eq:scalarproduct}
\IPROD{f}{g}\DEF\SPROD{f^+}{g^+}+
\sum_{i=0}^\infty\left\{
\SPROD{f}{\chi_i}\SPROD{\chi_i}{g}+\overline{f^i}g^i\right\},\quad 
\forall f,g\in\SK.
\end{equation}
We denote the Hilbert norm on $\SK$ by 
$\NORM{.}\DEF p(.)=\IPROD{.}{.}^{1/2}$.
In particular, we have the identity $\SPROD{f^+}{f^+}=\NORM{f^+}_{L^2}^2$.
\end{coro}
We set
\[
\SP\DEF\Bigl\{
f\in\SV\Bigm| f^{(k)}(0)=0,\ \forall k\in\NN_0
\Bigr\}.
\]
Obviously, \SPROD{.}{.} is positive definite on \SP and equals
the $L^2$-scalar product on that subspace.
The decomposition~\eqref{eq:decomposition} can now be expressed
as follows:
\begin{lemm}
\labelL{P}
Equation~\eqref{eq:decomposition} defines a mapping
\begin{displaymath}
  P\colon  \SV\longrightarrow\CLOT{\SV};\quad f\longmapsto f^+,
\end{displaymath}
with the following properties:
  \begin{enumerate}
    \item $P$ is continuous in the topology $\tau$.
    \item $P$ has a continuous extension to \SK.
    \item $P$ maps \SK onto \CLOT{\SP}.
    \item $P$ is an orthogonal projection onto \CLOT{\SP} with respect
      to \IPROD{.}{.}.
    \item The decomposition
    \[
      \SK=\CLOT{\SP}\oplus\CLOT{\SNN}
    \]
    is orthogonal with respect to the scalar product \IPROD{.}{.}
    (denoted by $\oplus$).
  \end{enumerate}
\end{lemm}
\begin{proof}
To show i) we estimate
\begin{align*}
  \NORM{f^+}^2 &=\NORM{f^+}_{L^2}^2 + 
     \sum_{i=0}^\infty \ABS{\SPROD{f^+}{\chi_i}}^2
   = \NORM{f^+}_{L^2}^2 + 
     \sum_{i=0}^\infty \ABS{\IPROD{f^+}{\chi_i}_{L^2}}^2\\
   &\leq \NORM{f^+}_{L^2}^2
    \left( 1+ \sum_{i=0}^\infty \NORM{\chi_i}_{L^2}^2\right)
   \leq \NORM{f^+}_{L^2}^2
    \left( 1+ \sum_{i=0}^\infty c_i^2\gamma_i^2\right)\\
   & \leq C \NORM{f^+}_{L^2}^2  \leq C \NORM{f}^2.
\end{align*}
In the last step we used that we have 
$\NORM{f}^2\geq \SPROD{f^+}{f^+} = \NORM{f^+}_{L^2}^2$
 by \refC{Hilbert}. Assertion ii) follows from i).
By i) and ii) it suffices to show that $f\in\SV$ entails
$f^+\in\CLOT{\SP}$ to show iii). For that, by the second to last
inequality mentioned previously, it suffices to approximate
$f^+$ in the $L^2$-norm with elements of \CLOT{\SP}.
Such an approximation can be easily constructed\eg as
$f^+_{\EPS}=(1-\rho_{\EPS})f^+$ for $\EPS\rightarrow0$, 
with the cut-off functions $\rho_{\EPS}$ of \refL{basic}. 
That $P\colon\SK\rightarrow\CLOT{\SP}$ is 
surjective is now clear, since $P$ is the identity on \CLOT{\SP}.
For $f^+\in\SP$ and $\chi\in\SNN$, the scalar product reduces to
\[
\IPROD{f^+}{\chi}=\sum_{i=0}^\infty 
  \SPROD{f^+}{\chi_i}\SPROD{\chi_i}{\chi}=0,
\]
since $\SPROD{\chi_i}{\chi}=0$, and because $f^+$ and $\chi$ have 
decompositions with vanishing $(f^+)^i$ and $\chi^+$, respectively.
Continuity of \SPROD{.}{.} then implies statement iv).
Assertion v) follows from iv) and the fact that the sum 
in~\eqref{eq:decomposition} converges to an element of \CLOT{\SNN}.
\end{proof}
To construct the metric operator $J$ that connects the indefinite 
with the Hilbert scalar product on \SK, we have to decompose this
space somewhat further. To that end, we consider the functionals
\[
F_i(f)\DEF\SPROD{\chi_i}{f},\quad f\in\SV,
\]
on \SV.
These functionals are nonzero since \SV is non-degenerate, they
vanish on \SNN, and they are clearly bounded with respect to the norm
$p$. In fact, we have $\ABS{F_i(f)}/p(f)\leq1$ for
$f\in\SV$, by~\eqref{eq:majorant}. That is, the $F_i$ have unique
continuations (also denoted by $F_i$) to \SK by the Hahn--Banach 
theorem, and by continuity these satisfy the same bound 
$0<\NORM{F_i}\leq 1$.
\begin{lemm}
  \labelL{repvectors}
The uniquely determined vectors $v_i\in\SK$ which represent
$F_i$ via $F_i(f)=\IPROD{v_i}{f}$ for all $f\in\SK$
are actually contained in \CLOT{\SP}.
\end{lemm}
\begin{proof}
  That the vectors $v_i$ exist and are unique in \SK follows
from Riesz' representation theorem applied to the bounded linear
functionals $F_i$ on the Hilbert space \SK. We have to show that
they are in \CLOT{\SP}. Choose a sequence $\{v_{in}\}_{n\in\NN}$
in \SV that approximates $v_i$\ie $\NORM{v_i-v_{in}}\rightarrow 0$
for $n\rightarrow\infty$. Using the decomposition~\eqref{eq:decomposition}
for the $v_{in}$ we calculate (adopting Einstein's summation convention
for repeated upper and lower indices)
%
\begin{alignat*}{3}
  \NORM{v_i-v_{in}}^2=&\ 
    \IPROD{v_i-(v_{in})^+-(v_{in})^j\chi_j}{v_i-(v_{in})^+-(v_{in})^k\chi_k} &\\
    =&\ \NORM{v_i-(v_{in})^+}^2 &\\
    & - \IPROD{v_i-(v_{in})^+-(v_{in})^j\chi_j}{(v_{in})^k\chi_k}
       - \IPROD{(v_{in})^j\chi_j}{v_i-(v_{in})^+} &\\
    =&\ \NORM{v_i-(v_{in})^+}^2 &\\
    & -  \IPROD{v_i-v_{in}}{(v_{in})^j\chi_j}
       - \IPROD{(v_{in})^j\chi_j}{v_i} +\IPROD{(v_{in})^j\chi_j}{(v_{in})^+}. &
\end{alignat*}
%
The last term on the right-hand side vanishes for all $n$ due
to \refL{P}v). The third term is zero since 
$\IPROD{v_i}{\chi_j}=F_i(\chi_j)=\SPROD{\chi_i}{\chi_j}=0$.
We use the Cauchy--Schwartz estimate for
the scalar product \IPROD{.}{.} and the fact that 
$
\NORM{\chi_i}=\ABS{(\chi_i)^i}=\ABS{\smash[t]{\chi_i^{(i)}(0)/\gamma_i}} =1
$
to estimate the second term as follows:
\[
\ABS{\IPROD{v_i-v_{in}}{(v_{in})^j\chi_j}}\leq
\NORM{v_i-v_{in}}\sum_{j=0}^\infty\ABS{\smash[t]{(v_{in})^j}}\leq C \NORM{v_i-v_{in}},
\]
with some constant $C>0$ independent of $n$. In fact, since $(v_{in})^j=
v_{in}^{(j)}(0)/\gamma_j$, and using~\eqref{eq:regularity} and~\eqref{eq:damping} 
we see that the sum is finite for all $n$. Since the sequence $v_{in}$ is
convergent in the norm $p$ and by definition~\eqref{eq:majorant} of this norm,
the sum must actually converge and therefore admits a global bound $C$ as above.
In conclusion, since  $v_{in}$ is $\tau$-convergent to $v_i$\ie 
$\NORM{v_i-v_{in}}\rightarrow 0$
for $n\rightarrow\infty$, we must have $\NORM{v_i-(v_{in})^+}\rightarrow 0$ 
by necessity, and thus already $Pv_{in}$ is $\tau$-convergent to $v_i$. 
This shows the claim.
\end{proof}
The basic properties of the $v_i$ are collected in the next lemma.
\begin{lemm}
  \labelL{vprop}
The vectors $v_i$ have the following properties:
\begin{enumerate}
\item $\SPROD{\chi_i}{v_i}=\IPROD{v_i}{v_i}=1$. 
\item $\IPROD{v_i}{v_j}=\SPROD{\chi_i}{v_j}=0$ for $i\neq j$.
\item $\SPROD{v_i}{v_i}=0$.
\item $\SPROD{v_i}{f}=f^i$ for all $f\in\SV$.
\end{enumerate}
\end{lemm}
\begin{proof}
Statement i) is clear from the defining property of $v_i$, except for
the last equality that says $\NORM{v_i}=1$. This will soon turn out to be true.
Let $\{v_{in}\}_{n\in\NN}\subset\SP$ be a sequence converging to $v_i$ in \SK, 
which exists by \refL{repvectors}. Then with~\eqref{eq:scalarproduct},
and since $(v_{in})^j=0$ for all $j$ we have
\[
\frac{\NORM{v_{in}}^2}{\ABS{\SPROD{\chi_i}{v_{in}}}}=
\frac{\SPROD{v_{in}}{v_{in}}}{\ABS{\SPROD{\chi_i}{v_{in}}}}+1+
\frac{\sum\limits_{j\neq i}\SPROD{\chi_j}{v_{in}}}{\ABS{\SPROD{\chi_i}{v_{in}}}}.
\]
Now $\ABS{\SPROD{\chi_i}{v_{in}}}=\ABS{\IPROD{v_i}{v_{in}}}\longrightarrow
\NORM{v_i}^2$ by i), so that the left-hand side tends to $1$ for 
$n\rightarrow\infty$ (here we assume that the denominators are
nonzero which can be achieved by choosing $v_{in}$ suitably). 
Since the denominators stay bounded, we must necessarily have 
$\ABS{\SPROD{\chi_j}{v_{in}}}
\longrightarrow 0$ for $j\neq i$ showing ii), 
and also $\SPROD{v_{in}}{v_{in}}\longrightarrow0$
showing iii) since $v_{in}$ converges to $v_i$ with respect to $\NORM{.}=p(.)$
and $p$ majorizes the inner square. 
Incidentally, this also shows
\[
\sup_n \frac{\ABS{F_i(v_{in})}}{\NORM{v_{in}}}=
\sup_n\frac{\ABS{\SPROD{\chi_i}{v_{in}}}}{\NORM{v_{in}}}=1,
\]
and this proves the last equality in i), since the norm of $v_i$ and that
of the linear functional $F_i$ coincide by Riesz' theorem.
To show iv), we consider again the 
decomposition~\eqref{eq:decomposition} of a vector $f\in\SV$ which yields
\[
\SPROD{v_{in}}{f}=\SPROD{v_{in}}{f^+}+f^i\SPROD{v_{in}}{\chi_i}+
\sum_{j\neq i}f^j \SPROD{v_{in}}{\chi_j}.
\]
In this expression we find $\SPROD{v_{in}}{f^+}\longrightarrow 0$, since
by iii) $v_{in}$ converges strongly to $0$ in \CLOT{\SP}, and due to \refL{P}.
By arguments similar to that in the proof of \refL{repvectors}, the
sum stays bounded independently of $n$, and since every single term in it
converges to $0$ by ii), the sum also tends to $0$. This leaves us with
the second term which converges to $f^i\SPROD{v_i}{\chi_i}=f^i$
by i). This shows iv).
\end{proof}
The $v_i$ could be constructed concretely as limits of functions which vanish
strongly in the $L^2$-sense, as in~\cite{MPS90}. \refL{repvectors} 
and~\ref{lemm:vprop} allow us to avoid such an explicit construction.
The vectors $v_i$ are an orthonormal basis of a closed Hilbert subspace
of \SK. This space is isomorphic to the dual space of \CLOT{\SNN} by definition
of the functionals $F_i$ and \refL{repvectors}, and we mnemonically
denote it by the symbol $\overline{\SNN}^{\tau\LIN{\ast}}$. 
\begin{lemm}
\labelL{Kdecomp}
  Denote by \SH the closure $\overline{\SP}^{\tau_+}$ of
\SP with respect to the topology $\tau_+$ induced by the quadratic Hilbert
norm $p_+(.)^2\DEF\SPROD{.}{.}$ on \SP. The space \SK admits the 
decomposition
\[
\SK=\SH\oplus\overline{\SNN}^{\tau\LIN{\ast}}\oplus \CLOT{\SNN},
\]
 orthogonal with respect to \IPROD{.}{.}.
\end{lemm}
\begin{proof}
First, we must show that the decomposition is indeed possible because
$\SH\subset\CLOT{\SV}$. To this end, note that the topology $\tau_+$ is
stronger than the restriction of $\tau$ to \SP. In fact, if a sequence in 
\SP converges
in the norm $p_+$ then it converges in the $L^2$-norm by~\eqref{eq:iprod},
and by the action of the indefinite product on \SP it is easy to see
that this suffices to ensure convergence in the norm $p$.
Now, taking~\refL{P}v) into account, we have to show that the 
\IPROD{.}{.}-orthogonal decomposition 
$\CLOT{\SP}=\SH\oplus\overline{\SNN}^{\tau\LIN{\ast}}$ holds.
First, observe that the vectors $v_i$ form a $\tau$-complete orthonormal
system in $\overline{\SNN}^{\tau\LIN{\ast}}$. Now, for
$f^+$, $g^+\in\SP$ we have 
\[
\IPROD{f^+}{g^+}=\SPROD{f^+}{g^+}+
\sum_{i=0}^\infty\IPROD{f^+}{v_i}\IPROD{v_i}{g^+},
\]
by the definition of $v_i$ and~\eqref{eq:scalarproduct}. 
This shows that a sequence $\{f^+_n\}_{n\in\NN}$ in \SP converges
to a limit $f\in\CLOT{\SP}$ if and only if $p_+(f_n^+-f)\longrightarrow 0$
and independently the \IPROD{.}{.}-orthogonal projection of $f^+_n-f$ onto
the closed subspace $\overline{\SNN}^{\tau\LIN{\ast}}$ of \CLOT{\SP}
tends to zero. Denote by $\overline{\SNN}^{\tau\LIN{\bot}}$ the
orthogonal complement of $\overline{\SNN}^{\tau\LIN{\ast}}$ in \SK
with respect to \IPROD{.}{.}. By the above-given argument, the subset
$\SP\cap\overline{\SNN}^{\tau\LIN{\bot}}$ of \CLOT{\SP} is dense in
\SH with respect to the topology $\tau_+$. This shows that
the proposed decomposition is indeed \IPROD{.}{.}-orthogonal.
In conclusion, a $\tau$-Cauchy sequence in \SP can be identified
with a pair $(f,\{\lambda_i\}_{i\in\NN_0})$ with an $f\in\SH$ and
$\lambda_i=\IPROD{v}{v_i}$ for some $v\in\overline{\SNN}^{\tau\LIN{\ast}}$.
This shows $\CLOT{\SP}=\SH\oplus\overline{\SNN}^{\tau\LIN{\ast}}$.
\end{proof}
It should be noted that by~\refL{vprop}iii), the vectors $v_i$ indeed converge 
to zero in the topology $\tau_+$ of \SH but are clearly nonzero in 
$\CLOT{\SP}\subset\SK$.
Furthermore, $\tau_+$ is stronger than the $L^2$-topology
although $p_+(f^+)=\NORM{f^+}_{L^2}$ for $f^+\in\SP$. We will characterize
\SH as a function space in the following. 
We have compiled all information needed to exhibit the Krein space structure 
of \SK.
\begin{theo}
\labelT{Hilbertstruct}
  The space \SK is a Krein space with countably infinite rank of indefiniteness.
Its Hilbert space structure is maximal and given by the metric operator 
$J\colon \SK\rightarrow\SK$, satisfying
$\SPROD{.}{.}=\IPROD{.}{J.}$.  It holds
\[
J v_i=\chi_i,\ J\chi_i=v_i\text{, and }J|_{\SH}=\mathbb{I}_{\SH},
\]
in the decomposition of \refL{Kdecomp}.
\end{theo}
\begin{proof}
  The strategy of the proof will be as follows: 
The metric operator exists by~\refP{metricop}, and 
we have seen in~\refL{Kdecomp} that we can write down its action
in the decomposition 
$\SK=\SH\oplus\overline{\SNN}^{\tau\LIN{\ast}}\oplus \CLOT{\SNN}$.
We can then explicitly demonstrate that  the operator $J$ on \SK acts as
stated. This special form of $J$  immediately implies that it is a bounded, 
completely invertible operator on \SK. Thus by~\refP{Krein}, 
\SK is a Krein space and since $J^{-1}=J$ is also bounded, its Hilbert 
space structure $(\SK,J)$ is maximal by~\refL{maximal}.
Now, by definition of the $v_i$ we have 
$\SPROD{f}{\chi_i}=\IPROD{f}{J\chi_i}=\IPROD{f}{v_i}$ for all $f\in\SK$,
showing $J\chi_i=v_i$. On the other hand, by~\refL{vprop}iv) 
and~\eqref{eq:scalarproduct} we have $\SPROD{f}{v_i}=
\IPROD{f}{Jv_i}=\overline{f^i}=
\IPROD{f}{\chi_i}$, showing $Jv_i=\chi_i$.
It remains to consider the restriction of $J$ to \SH. Take
$f^\bot$, $g^\bot\in\SP\cap\overline{\SNN}^{\tau\LIN{\bot}}$ (see the proof
of~\refL{Kdecomp}) and note that 
$\IPROD{\smash{f^\bot}}{\smash{g^\bot}}=\SPROD{\smash{f^\bot}}{\smash{g^\bot}}$
for those vectors. Since these vectors are dense in \SH, it follows that the
restriction of $J$ to \SH is the identity. This shows the claim.
\end{proof}
To conclude the proof of~\refT{main}, it finally remains to show that \SH 
is the Fourier transform of the  space $L_0^2$ defined in \refD{Lnull}.
Now \FOUR is a topological isomorphism from \SUD{\alpha}{\beta} onto
$\SV=\SUD{\beta}{\alpha}$ and for $f\in\SUD{\alpha}{\beta}$ we have
\[
  i^k\widehat{f}^{(k)}(0)=
    \left(i^k\frac{\dd^k}{\dd\xi^k}\int_\RR \ee^{-ix\xi}f(x)\dd x
    \right)\Bigg|_{\xi=0}
  = i^k\int_\RR (-ix)^k f(x) \dd x = \MOM[f]{k}.
\]
By that, the image of \SP under $\FOUR^{-1}$ is the subspace of 
\SUD{\alpha}{\beta} of functions $f$ with $\MOM[f]{k}=0$ for all $k$.
Since \SUD{\alpha}{\beta} is dense in $L^2$ and the Fourier transformation
is an $L^2$-isometry, we can see $\SH=\FOUR L_0^2$. Thus~\refT{main}
is finally proven.

If we test the vectors $v_i$ with states in the `physical' subspace \SH\ie
the representation space for the Heisenberg-observables, they appear as
completely delocalized states. In fact the action of the momentum operator
on them is given by \refL{vprop}iv):
\[
\SPROD{pv_i}{f}=\SPROD{v_i}{pf}=(pf)^i=0,
\quad\text{for all }f\in\SH,\ i\in\NN_0,
\]
where we denoted the unique extension of the multiplication operator $p$
from \SV to \SK also by $p$.
This is different from the case of one single negative degree of freedom 
in~\cite{MPS90}, where the
single vector $v_0$ turns out to be completely delocalized on the whole Krein
state space.
\section{A Condition Sufficient for Regularization}
\labelS{generalized}
In this last section, we want to give a set of conditions on a general 
indefinite inner product space \SV, that will be sufficient for the
regularization procedure to work. We did not put this generalization in the 
beginning, and then deduced the special case $\SV=\SUD{\beta}{\alpha}$ considered
previously from it, for two reasons: First and foremost, we wanted to 
emphasize the case of indefinite inner products generated by singular kernels
acting on a test function space, which we think is particularly interesting
in view of possible applications in physics. Second,
most of the assertions and proofs in \refS{regularization} are already cast 
abstract enough to be re-used in the proof of the generalized regularization
\refT{generalized} without any modification. Thus, we can stress the  
essential points that need modification and thereby highlight the principles
which put the regularization procedure to work.

Two elements are essential: First, the existence of neutral decomposition
elements $\chi_i$ that enable us to isolate the positive part of the indefinite
product. Second, a certain balance between the growth, respectively, and decay
of a) the inner products
of vectors in the space with the neutral elements, and b) the coefficients of the 
linear decomposition of a vector with respect to these. These growth 
conditions constitute the main difference between the case of finite rank
of indefiniteness considered in~\cite{SCH97A} and the infinite case, where
they serve to render the Hilbert majorant topology well-defined in the first 
place. 

Let us now formulate our set of conditions. We assume \SV to be a complex linear
space with an indefinite inner product \SPROD{.}{.}, which shall be non-degenerate. 
Assume that:
\begin{enumerate}
\renewcommand{\labelenumi}{\arabic{enumi})}
\setcounter{enumi}{-1}
\item There exists an orthogonal system $\{\widetilde{\chi}_i\}_{i\in{\NN_0}}$ 
of mutually linearly independent, neutral vectors in \SV.
\item For all $v\in\SV$, the unique decomposition
for $N\in\NN_0$,
\[
v=\widetilde{v}^{N+}+\sum_{i=0}^N \widetilde{v}^i\widetilde{\chi}_i,
\quad \widetilde{v}^i\in\CC,
\]
becomes asymptotically positive in the sense that
\[
0\leq\lim_{N\rightarrow\infty}
\SPROD{\smash{\widetilde{v}^{N+}}}{\smash{\widetilde{v}^{N+}}}.
\]
\item There exists a sequence of complex numbers $\{\gamma_i\}_{i\in\NN_0}$
such that both sequences $\{ \gamma_i\SPROD{\widetilde{\chi}_i}{v}  \}$ 
and $\{ \widetilde{v}^i/\gamma_i \}$ are in $l^2(\NN_0)$. 
\end{enumerate}
These conditions enable us to prove an equivalent of \refP{majorant}. In fact,
setting $\chi_i\DEF\gamma_i\widetilde{\chi}_i$, we obtain the anologue of the
finite decomposition~\eqref{eq:finite_decomp} for a vector $v\in\SV$ 
with coefficients $v^i=\widetilde{v}^i/\gamma_i$. We then have to see that
the sum~\eqref{eq:majorant} with $f$ replaced by $v$, defining the majorant norm 
$p(v)^2$, converges. The convergence of the asymptotically positive part
$\lim_{N\rightarrow\infty}\SPROD{\smash{v^{N+}}}{\smash{v^{N+}}}$ 
then follows, as we have already noted after the proof of \refP{majorant} 
on page~\pageref{conv_note}. Now, the $i$th summand in the definition of
$p(v)^2$ becomes
\[
  \ABS{\SPROD{v}{\chi_i}}^2+\ABS{\smash{v^i}}^2 = 
  \ABS{\gamma_i\SPROD{v}{\widetilde{\chi}_i}}^2+
  \ABS{\smash{\widetilde{v}^i/\gamma_i}}^2, 
\]
and the sum converges due to condition 2). Thus, we get a majorant Hilbert topology $\tau$
on \SV. A close inspection of the proofs
of the various lemmata in Section~\ref{sect:regularization} shows that the only
other point which has to be reconsidered is the proof of statement
i) of \refL{P}, that the mapping $P\colon v\mapsto v^+$ is $\tau$-continuous on
$\SK\DEF\CLOT{\SV}$. There, we have utilized the $L^2$-norm, but we
will see that this can also be shown independently. In fact, we have
\begin{align*}
  \NORM{\smash{v^+}}^2&=\IPROD{v-\smash[t]{\sum_{i=0}^\infty} v^i\chi_i}{v-\smash{\sum_{j=0}^\infty} v^j\chi_j}\\
     &= \NORM{v}^2 - 2\RE \sum_{i=0}^\infty v^i \IPROD{\chi_i}{v} +
        \sum_{i,j=0}^\infty v^i\overline{v^j}\IPROD{\chi_i}{\chi_j}.\\
\intertext{We use the two consequences $\IPROD{\chi_i}{\chi_j}=\delta_{ij}$ 
and $\IPROD{\chi_i}{v}=\overline{v^i}$ of \refE{scalarproduct} in the third and
second term respectively to obtain}
     &= \NORM{v}^2 -  \sum_{i=0}^\infty \ABS{\smash{v^i}}^2 \leq \NORM{v}^2, 
\end{align*}
by definition~\eqref{eq:majorant} of $\NORM{v}^2$. Here again, condition~2) ensures
the convergence of the sums appearing. From this point, one can proceed word for word as
in \refS{regularization} with the definition of the vectors $v_i$ and the demonstration
of their properties. We finally obtain a generalization of \refT{Hilbertstruct}:
\begin{theo}\labelT{generalized}
Let \SV satisfy~0)--2). Then
\SK is a Krein space with rank of indefiniteness equal to
$\#\{\chi_i\neq 0\}$. Its Hilbert space structure is maximal, and the
metric operator $J$ acts as in \refT{Hilbertstruct}.
\end{theo}
Note that~0)--2) and \refT{generalized} are formulated as to cover the
cases of finite as well as of infinite rank of indefiniteness. Namely, in case
the rank of indefiniteness is $N<\infty$, one can find at most $N$ neutral, linearly
independent vectors, and one has to use them all to obtain a decomposition that, as
demanded by~1), becomes positive (in this case not asymptotically). One then
chooses $\widetilde{\chi}_i=0$ for $i>N-1$.

We conclude this paper with some comments on the generalized regularization procedure
just described. First, the conditions~0)--2) certainly do not present the
utmost general ones possible. In particular, one can perhaps  replace the
neutral orthogonal system of~0) by a general system of linear independent vectors
which lead to an aymptotically positive decomposition. See~\cite[Remark~A.13]{SCH97A},
where we describe how to find a maximal neutral orthogonal system in the case of finite
rank of negativity. Furthermore, whether
$\SPROD{\smash{v^+}}{\smash{v^+}}$ is positive or negative definite is irrelevant, 
since one can always go over to $-\SPROD{.}{.}$ (the so called \emph{anti-space} 
of \SV). On the other hand, one cannot easily dispense with either of the
$l^2$-conditions in~2), since they represent rather sharply the convergence conditions
that enabled us to construct a majorant.  Since we made no presuppositions with
respect to \SV regarding structure and topology, condition~1) is also indispensible.

In our case of main interest in \refS{regularization}, the essence of conditions~0) 
and~1) are captured in \refL{decomposing_functions} which is proven in the appendix
following this section. 
A similar construction of neutral decomposition elements will also have to be
carried out in any other concrete case, and is thus at the very center of the
regularization procedure, in putting flesh to the bones of the abstract
conditions~0) and~1). The construction in Appendix~\ref{sect:neutral} 
may serve as a blueprint for that at least in the case of test function spaces 
over $\RR^n$ and inner products generated by kernels whose singularities are `localized' 
enough\eg concentrated on a compact set. This may justify that we did not delve into a 
further abstraction of conditions~0) and~1). 

Let us consider an instructive special case. Assume the sequence of
coefficients $\{ \widetilde{v}^i \}$ is bounded for all $v\in\SV$. 
If there holds an estimate
\[
\ABS{\SPROD{\widetilde{\chi}_i}{v}}\leq C(v) i^{-(1+\delta)},
\]
with a constant depending on $v$ and for some $\delta>0$, we can choose
\[
\gamma_i=i^{-(1/2+\EPS)},
\]
for any $0<\EPS<\delta$. Such polynomial growth and decay conditions are obviously
much weaker than the conditions that were present in the case $\SV=\SUD{\beta}{\alpha}$,
see\eg our choice of $\gamma$'s in~\eqref{eq:damping}. Thus the range of cases covered
by \refT{generalized} is considerably widened in comparison to \refT{Hilbertstruct}
and \refT{main}.

The question arises naturally, whether we can find uniform properties on \SV, 
as opposed to the pointwise ones 1) and 2), that enable regularization. In essence
one would look for a simple quantitative measure that tells us when the construction
of the majorant is possible. But this is not straightforward. To simplify the
discussion, consider the case where $\{\widetilde{v}^i \}$ is bounded in \CC for
all $v$ (these sets can of course not be uniformly bounded). A simple uniform measure
that could replace condition 2) can be formulated in terms of the quantities
\[
\widetilde{\beta}_i\DEF 
  \sup\limits_{v\in\SV,\ \widetilde{v}^i=1}\ABS{\SPROD{\widetilde{\chi}_i}{v}}.
\]
Notice that at least $\widetilde{\chi}_i$ is in the set over which the supremum is
taken, and if this is the only element we have $\widetilde{\beta}_i=0$ due to 
neutrality of that vector. One can then replace 2) by the condition that there
shall exist a sequence $\{\gamma_i^{-1}\}$ in $l^2(\NN_0)$ such that also
$\{\widetilde{\beta}_i\gamma_i\}$ is in $l^2(\NN_0)$. This uniform growth
condition on $\widetilde{\beta}_i$ is however by far too restrictive, since
it does not even cover the case considered in \refT{main}. The basic reason for this
is that in most cases the neutral orthogonal system $\{\widetilde{\chi}_i\}$ 
does not exhaust the space \SV in the sense that a complete orthogonal system
exhausts a Hilbert space. The inner products with these vectors do
not contain enough information about the whole space, and especially its positive
part, to decide sharply whether \SV is regularizable. The problem of finding
a good abstract definition of what we would like to call `regularizable inner
product spaces' remains therefore open.  
\appendix
\section{Construction of Neutral Decomposition Elements}
\label{sect:neutral}
In this Appendix, we present a simple construction for the 
neutral decomposing functions of~\refL{decomposing_functions}. 
We point out that
different and more refined constructions are surely possible, but
the one given in the following suffices for our purpose.

We have to show i)--iii) since iv) follows from them.  We
prove~\refL{decomposing_functions} for $\gamma_k=1,~\forall k$. 
The general case follows by multiplication of the functions $\chi_k$ 
constructed subsequently with the given sequence $\gamma_k$. The first thing 
we need to show is that there are enough functions of compact support 
in \SUD{\beta}{\alpha}. For that, we have to consider the spaces
\SUD{\beta,B}{\alpha}, which constitute \SUD{\beta}{\alpha} as an
inductive limit for $B\rightarrow\infty$, 
see~\cite[Chapter~IV,~\S3]{b:GS64} for their definition.
\begin{lemm}
  \labelL{basic} 
  Let $0\leq\alpha\leq\infty$ and $1<\beta<\infty$. For
  $\EPS>0$ there exists $B_{\EPS}>0$ and a function
  $\rho_{\EPS}\in\SUD{\beta,B_{\EPS}}{\alpha}$ such that
\[
\rho_{\EPS}(x)=
  \begin{cases}
    1, &\text{if $\ABS{x}<\EPS/2$;}\\
    0, &\text{if $\ABS{x}>3\EPS/2$;}\\
    0\leq\rho_{\EPS}(x)\leq1, &\text{otherwise.}
  \end{cases}
\]
\end{lemm}
\begin{proof}
  Under the given conditions, the function $\rho_{\EPS}$ can be
  constructed using the well-known facts about the Gelfand--Shilov
  spaces, for which we refer to~\cite[Chapter~IV]{b:GS64}. The space
  \SUD{\beta}{\alpha} contains the space \SUD{\beta}{0} which consists
  of functions of compact support and is nontrivial for $\beta>1$.
  Furthermore, for $\phi\in\SUD{\beta}{0}$ we have
  $\phi^2\in\SUD{\beta}{0}$. Thus there exists a $B>0$ and a nonzero
  function $\phi$ with $\phi(x)\geq 0$ in \SUD{\beta,B}{0}, such that
  $\supp\phi\subset[-R,R]$ for some $R>0$. Then
\[
\phi_{\EPS}(x)\DEF \frac{\EPS}{2R\NORM{\phi}_{L^1}}\cdot\phi(2Rx/\EPS)
\]
is an element of \SUD{\beta,B_{\EPS}}{0} for $B_{\EPS}=2RB/\EPS$,
see~\cite[p.~158]{b:GS64}. It has $L^1$-norm $1$ and support in
$[-\EPS/2,\EPS/2]$. Since convolution with $L^1$-functions does not
change the regularity, the function $\phi_{\EPS}\ast\chi_{[-\EPS,\EPS]}$
is an element of \SUD{\beta,B_{\EPS}}{0} and therefore \textit{a
  fortiori} of \SUD{\beta,B_{\EPS}}{\alpha} with all the desired
properties.
\end{proof}
We set
\[
\kappa_n(x)\DEF \rho_{1/3}(x-n)/\NORM{\smash{\rho_{1/3}}}_{L^2},\quad n\in\ZZ.
\]
Since $\kappa_n$ has support in $[n-1/2,n+1/2]$, we have
$\IPROD{\kappa_i}{\kappa_j}=0$ for all $i\neq j$.  Define
a sequence of positive real numbers by
\begin{equation}
  \label{eq:eps_def}\tag{\dag}
  \EPS_i\DEF\frac{1}{3\ee}\prod_{k=0}^i \min(1,c_k^2).
\end{equation}
Set
\[
\delta_i\DEF\dfrac{x^i}{i!}\rho_{\EPS_i}.
\]
Furthermore, for $i\neq j$ define
\[
k_{ij}\DEF
\sign(i-j)\sqrt{\smash[b]{\IPROD{\delta_j}{\delta_i}_{L^2}}}.
\]
We use the following enumeration for the off-diagonal
entries of an infinite matrix (rows and columns counted from $0$):
\[
\NN_0^2\setminus\diag\ni(i,j)\longmapsto n(i,j)\DEF
\begin{cases}
  \frac{j(j-1)}{2}+i+1, &\text{if }i<j;\\
  n(j,i), &\text{otherwise.}
\end{cases}
\]
We use the functions $\delta_i$ as building blocks for the desired
functions, since they have just the right behaviour at $0$ to ensure
property ii) of~\refL{decomposing_functions}. To correct their nonvanishing
$L^2$-overlap with each other we use the corrective
\[
K_i\DEF\sum_{j\neq i}k_{ij}\kappa_{n(i,j)}.
\]
We must show that this is possible\ie that
$\NORM{\delta_i+K_i}_{L^2}^2$ does not exceed $c_i^2$, in order to
satisfy i). We have
\[
\NORM{\delta_i+K_i}_{L^2}^2=\NORM{\delta_i}_{L^2}^2+\NORM{K_i}_{L^2}^2=
\sum_{j=0}^\infty\ABS{\IPROD{\delta_i}{\delta_j}_{L^2}}.
\]
The terms in the sum allow for the basic (yet very coarse) estimate
\begin{displaymath}
  \ABS{\IPROD{\delta_i}{\delta_j}_{L^2}}\leq
  \frac{2}{i!j!}\left(\frac 3 2 \min(\EPS_i,\EPS_j)\right)^{i+j+1}
\end{displaymath}
by construction of $\delta_i$. Using~\eqref{eq:eps_def} we have
$\min(\EPS_i,\EPS_j)\leq\frac{1}{3\ee}\min(c_i^2,1)$ and therefore
we can estimate under the additional assumption $c_i^2\leq1$:
\begin{align*}
  \NORM{\delta_i+K_i}_{L^2}^2 & \leq \sum_{j=0}^\infty
  \frac{2}{i!j!}\left(\frac{c_i^2}{2\ee} \right)^{i+j+1}
  \leq\frac{2}{i!}\left(\frac{c_i^2}{2\ee}\right)^{i+1}
  \sum_{j=0}^\infty
  \frac{c_i^{2j}}{j!}\\
  &=\frac{2}{i!}\left(\frac{c_i^2}{2\ee}\right)^{i+1}\ee^{c_i^2} \leq
  \frac{c_i^2}{i!(2\ee)^i}\leq c_i^2.
\end{align*}
Now using the function
\[
\nu_i=\sqrt{c_i^2-\NORM{\delta_i+K_i}_{L^2}^2}\cdot\kappa_{-i}
\]
we can set
\[
\chi_i\DEF \delta_i+K_i+\nu_i.
\]
We are done if we show that $\chi_i$ is well defined as a function in
\SUD{\beta}{\alpha}\ie that the sum $K_i$ converges in the topology of
the namely space. To that end, we have to show convergence in one of the
spaces \SUD{\beta,B}{\alpha,A} which constitute the inductive limit
$\SUD{\beta}{\alpha}= \varinjlim_{A,B\rightarrow\infty}
\SUD{\beta,B}{\alpha,A}$.  We show that the increments in the sum
$K_i$, namely $k_{ij}\kappa_{n(i,j)}$, decay fast enough in $j$ to
turn the sum into a Cauchy sequence in that topology. In the countably
normed space \SUD{\beta,B}{\alpha,A}, we therefore have to estimate
the increments due to the terms in the sum defining $K_i$:
\[
\NORM{\smash{k_{ij}\kappa_{n(i,j)}}}_{\rho,\delta}= \sup_{x,k,q}
\frac{\ABS{\smash{x^kk_{ij}\kappa_{n(i,j)}}}}%
{(A+\rho)^k(B+\delta)^qk^{k\alpha}q^{q\beta}},\quad
\text{with }\rho,\delta>0.
\]
We first argue that this quantity can be estimated independently of
$q$. In fact, the functions $\kappa_{n(i,j)}$ are translates of a
fixed function in $\SUD{\beta,B_{1/3}}{\alpha}$, and therefore the supremum
over $q$ is smaller than a constant times the supremum over $k$ and
$x$ only, if we choose $B=\max(B_{1/3},B_{\EPS_i})$:
\begin{align*}
  \NORM{\smash{k_{ij}\kappa_{n(i,j)}}}_{\rho,\delta}&\leq C_\kappa
  \sup_{x,k}
  \frac{\ABS{\smash{x^kk_{ij}\kappa_{n(i,j)}}}}%
  {(A+\rho)^kk^{k\alpha}}.\\
  \intertext{It suffices to consider this especially for $A\geq1$ and
    $\alpha=1$ in which case we have} &\leq C_\kappa \sup_{x,k}
  \frac{\ABS{\smash{x^kk_{ij}\kappa_{n(i,j)}}}}{k^k}.  \intertext{For
    $j$ large enough and by definition of $n(i,j)$ we can estimate
    $\ABS{x}\leq2j^2$  on the support of $\kappa_{n(i,j)}$, and 
    with some other constant $C_\kappa'$
    depending only on the function $\kappa_{n(i,j)}$,} 
&\leq C_\kappa' k_{ij} \sup_k
  \left(\frac{2j^2}{k}\right)^k.  \intertext{Continuous maximization
    in $k$ shows} &\leq C_\kappa' k_{ij} \ee^{\displaystyle{cj^2}}.
\end{align*}
Now it is clear from their definition that $k_{ij}$ decay faster than
an exponential function of any type in $j$ and thus the claim follows.
\section{Basics of Indefinite Inner Product Spaces}
\label{sect:indef}
In this section we recall some facts about indefinite inner product, 
Krein and Pontryagin spaces needed in the main text.
For an extensive discussion of the subject matter we refer the
reader to~\cite{b:BOG74,b:AI89}. First some notations: 
Let $\SV$ be a vector space equipped with an indefinite inner product
$\SPROD{.}{.}$
(antilinear in the first, linear in the second argument). 
\newcommand{\SA}{\ensuremath{\mathord{\mathcal{A}}}\xspace}
The linear span of a subset \SA of vectors in \SV
is denoted by \LIN{\SA}. The \emph{linear sum} of subspaces 
$\SV_1,\ldots,\SV_n$ of \SV is given by $\LIN{\SV_1\cup\cdots\cup\SV_n}$
and denoted by $\SV_1+\cdots+\SV_n$. If the spaces $\SV_1,\ldots,\SV_n$
are linearly independent, their linear sum is termed \emph{direct sum}
and denoted by $\SV_1\dotplus\cdots\dotplus\SV_n$. \emph{Orthogonality} with
respect to
$\SPROD{.}{.}$ is defined, and denoted by the binary relation $\bot$ as usual
(but clearly does not have the same strong consequences as in definite
inner product spaces). If the $\SV_1,\ldots,\SV_n$ are mutually orthogonal,
\newcommand{\os}{\mathbin{(\dotplus)}}
their \emph{orthogonal direct sum} is denoted by $\SV_1\os\cdots\os\SV_n$,
whereas the symbol $\oplus$ is reserved for orthogonal sums with respect to a 
positive definite inner product, which we will denote with
$\IPROD{.}{.}$, following mathematical convention. By \emph{positive definite} 
we mean as usual
$\IPROD x x \geq 0$, $\forall x \neq 0$, and $\IPROD x x =0 \Rightarrow x=0$.
A subspace \SA of \SV is called \emph{positve}, 
\emph{negative}, or \emph{neutral}, respectively, if one of the possibilities
$\SPROD x x > 0$, $\SPROD x x <0$ or $\SPROD x x =0$ holds for all 
$x\in\SA$, with $x\neq 0$. One sets
$$
\SV^{++}\DEF\bigl\{x\in\SV\bigm|\,\SPROD x x > 0\text{ or } x=0\bigr\},
$$
and calls this subset the \emph{positive part} of \SV. The \emph{negative}
and \emph{neutral} parts $\SV^{--}$ and $\SV^0$ are defined alike.
A subspace \SA of \SV is called \emph{degenerate}, if its
\emph{isotropic part} $\SA \cap \SA^\bot$ does not only consist
of the zero vector. In the main text and the 
following we will deal merely with \emph{non-degenerate} spaces\ie
spaces with $\SV^\bot=\{0\}$.
A non-degenerate inner product space \SV is said to be \emph{decomposable} 
if it admits a \emph{fundamental decomposition}
$$
\SV=\SV^\bot\os\SV^+\os\SV^-,\quad \text{with } \SV^+\subset\SV^{++},\ 
\SV^-\subset\SV^{--}.
$$
For non-degenerate spaces the isotropic part of the decomposition vanishes.
The dimension of a maximal negative definite subspace $\SV^-\subset\SV^{--}$
appearing in a fundamental decomposition of a non-degenerate inner product 
space is called the \emph{rank of negativity} of \SV. 
As proven in \cite[Corrollaries~II.10.4 and~IV.7.4]{b:BOG74},
it is an unique positive cardinal denoted by $\varkappa^-(\SV)$. 
The \emph{rank of positivity} $\varkappa^+(\SV)$ is defined in analogy to that. 
We set $\varkappa\equiv\min(\varkappa^-,\varkappa^+)$ and call 
this number the \emph{rank of indefiniteness} of \SV. 

Now some less trivial things about the topology of indefinite inner product 
spaces: A locally convex topology $\tau$ on \SV defined by a single 
seminorm $p$, which is then actually a norm, is called \emph{normed}.
If \SV is $\tau$-complete, we say that $\tau$ is a 
\emph{Banach topology}. If
$\tau$ can be defined by a \emph{quadratic norm} $p(x)=\IPROD{x}{x}^{1/2}$,
where $\IPROD{.}{.}$ is a positive definite inner product on \SV, 
then $\tau$ is called
a \emph{quadratic normed topology}. Again, if \SV is $\tau$-complete,  
then $\tau$ is termed
\emph{Hilbert topology}. A normed topology $\tau_1$ is \emph{stronger}
than another $\tau_2$, written $\tau_1\geq\tau_2$, if and only if every 
$\tau_2$-open set is also a $\tau_1$-open set, or equivalently the relation 
$p_1(x)\geq\alpha p_2(x)$ holds for all $x\in\SV$, with an $\alpha>0$. Two
norms that define the same topology are called \emph{equivalent}.
A locally convex topology $\tau$ on $\SV$ is called a \emph{partial majorant}
of the inner product if $\SPROD{.}{..}$ is separately $\tau$-continuous. 
The \emph{weak topology} on $\SV$ is the topology defined by the family of 
seminorms
\[
p_y(x)\DEF\ABS{\SPROD{y}{x}},\quad\text{for all }x\in\SV.
\]
\begin{lemm}[{\cite[Theorem II.2.1]{b:BOG74}}]
The weak topology is the weakest partial majorant on \SV.
If a locally convex topology on $\SV$ is stronger than the weak topology,
then it is a partial majorant.
\end{lemm}
We will need a stronger concept of topology:
\begin{defi}
\labelD{majorant}
A locally convex topology $\tau$ on $\SV$ is called \emph{majorant topology}, if
the inner product $\SPROD{.}{..}$ is jointly $\tau$--continuous. 
\end{defi}
In applications, one can often restrict oneself
to majorants defined by a single seminorm which majorizes the
inner square, as we can see from the following result.
\begin{lemm}[{\cite[Lemma IV.1.1 \& 1.2]{b:BOG74}}]
\labelL{norm_maj}
It holds:
\begin{enumerate}
  \item To every majorant there exists a weaker majorant defined by
    a single seminorm.
  \item For a locally convex topology defined by a single seminorm
    $p$ to be a majorant it is sufficient that $p$ dominates the inner
    square:
    $$
    \ABS{\SPROD x x} \leq \alpha p(x)^2,\quad\alpha>0,\ \forall x\in\SV.
    $$
\end{enumerate} 
\end{lemm}
Majorant topologies, and especially majorant Hilbert topologies, have
many advantages over partial majorants. Before we describe them, let us see
why one would not like to use the weak topology on  general
indefinite inner product spaces:
\begin{lemm}[{\cite[Theorem IV.1.4]{b:BOG74}}]
The weak topology on the non-degenerate indefinite inner product space $\SV$
is a majorant, if and only if $\dim \SV<\infty$.
\end{lemm}
The indefinite inner product on a space  equipped with a majorant
Hilbert topology admits a simple description by the so-called \emph{metric
operator}.
\begin{prop}[{\cite[Theorem IV.5.2]{b:BOG74}}]
\labelP{metricop}
Let $\SV$ be an indefinite inner product space with a majorant Hilbert
topology $\tau$ defined by a norm $\NORM{.}$. Then there exists a
Hermitean linear operator, called \textrm{metric} (or \textrm{Gram}) operator,
$J$ on $\SV$ such that 
$$
\SPROD x y = \IPROD x {Jy},\quad \forall x,y\in\SV,
$$
where $\IPROD . .$ is the positive inner product on $\SV$ that defines
$\NORM{.}$. Moreover, in this case $\SV$ is decomposable and the fundamental
decomposition can be chosen so that each of the three components is 
$\tau$-closed.
\end{prop}
The spaces we want to construct in the main text should be complete in a 
certain sense, which we will now specify.
\begin{defi}
If a non-degenerate indefinite inner product space $\SK$ admits a decomposition
$$
\SK=\SK^+\os\SK^-,\quad \SK^+\subset\SK^{++},\ \SK^-\subset\SK^{--},
$$
such that $\SK^+$, $\SK^-$ are complete with respect to the restrictions of
the weak topology to them (termed \emph{intrinsically complete}),
then the space $\SK$ is called a \emph{Krein space}.
\end{defi}
Krein spaces can easily be characterized:
\begin{prop}[{\cite[Theorem V.1.3]{b:BOG74}}]
\labelP{Krein}
An indefinite inner product space $\SV$ is a Krein space if and only if
there exists a majorant Hilbert topology $\tau$ on $\SV$ such that
 metric operator $J$ is completely invertible.
\end{prop}
A Hilbert-space completion $\SH$ of an indefinite inner 
product space $\SV$, if 
it exists together with its metric operator $J$, is called
the \emph{Hilbert space structure} $(\SH,J)$ associated to $\SV$.
In applications one would like to find the largest Hilbert space associated to 
an indefinite inner product space. For that, one considers \emph{minimal}
majorant topologies\ie topologies $\tau_\ast$ such that no majorant
$\tau$ is weaker than $\tau_\ast$. Hilbert space structures given by
the completion of $\SV$ with respect to a minimal majorant are correspondingly
called \emph{maximal}. We find that the Hilbert space structure is maximal,
if  it leads actually to a Krein space:
\begin{lemm}[{\cite[Appendix A.1]{b:STR93}}]
\labelL{maximal}
A majorant Hilbert topology leads to a maximal Hilbert space structure
$(\SK,J)$, if and only if $J$ has a bounded inverse. 
Given  a Hilbert space structure
one can always construct a maximal one.
\end{lemm}
The last statement means in effect that every space admitting some majorant 
Hilbert topology can be completed to a Krein space.
\newcommand{\noopsort}[1]{} \newcommand{\singleletter}[1]{#1}
\providecommand{\bysame}{\leavevmode\hbox to3em{\hrulefill}\thinspace}
\providecommand{\MR}{\relax\ifhmode\unskip\space\fi MR }
\providecommand{\MRhref}[2]{%
  \href{http://www.ams.org/mathscinet-getitem?mr=#1}{#2}
}
\providecommand{\href}[2]{#2}

\end{document}